\documentclass[10pt]{article}
\usepackage[left=1.25in, right=1.25in, top=1.25in, bottom=1.25in]{geometry}
\usepackage[utf8]{inputenc}

\usepackage{cite}

\usepackage{nameref,hyperref}

\usepackage{microtype}

\usepackage{amsmath}
\DisableLigatures[f]{encoding = *, family = * }

\raggedright
\usepackage{changepage}

\usepackage[aboveskip=1pt,labelfont=bf,labelsep=period,singlelinecheck=off]{caption}

\makeatletter
\renewcommand{\@biblabel}[1]{\quad#1.}
\makeatother

\usepackage{color}

\definecolor{Gray}{gray}{.25}

\usepackage{graphicx}

\usepackage{sidecap}

\usepackage{wrapfig}
\usepackage[pscoord]{eso-pic}
\usepackage[fulladjust]{marginnote}
\reversemarginpar

\begin{document}
\vspace*{0.35in}

\begin{flushleft}
{\Large
\bf\textbf\newline{A computational approach to calculate the heat of transport of aqueous solutions}
}
\newline
\\
Silvia Di Lecce\textsuperscript{1},
Tim Albrecht\textsuperscript{1},
Fernando Bresme\textsuperscript{1,*}
\\
\bigskip
\textit{{1} Department of Chemistry, Imperial College London, SW7 2AZ, United Kingdom}
\\
\bigskip
* f.bresme@imperial.ac.uk

\end{flushleft}
\vspace*{0.35in}
\justify
\section*{Abstract}
Thermal gradients induce concentration gradients in alkali halide solutions, and the salt migrates towards hot or cold regions depending on the average temperature of the solution. This effect has been interpreted using the heat of transport, which provides a route to rationalize thermophoretic phenomena.
Early theories provide estimates of the heat of transport at infinite dilution. These values are used to interpret thermodiffusion (Soret) and thermoelectric (Seebeck) effects. However, accessing heats of transport of individual ions at finite concentration remains an outstanding question both theoretically and experimentally. Here we discuss a computational approach to calculate heats of transport of aqueous solutions at finite concentrations, and apply our method to study lithium chloride solutions at concentrations $>0.5$~M. The  heats of transport are significantly different for Li$^+$ and Cl$^-$ ions, unlike what is expected at infinite dilution.  We find  theoretical evidence for the
existence of minima in the Soret coefficient of LiCl, where the magnitude of the heat of transport is maximized. 
The Seebeck coefficient obtained from the ionic heats of transport varies significantly with temperature and concentration. We identify thermodynamic conditions leading to a maximization of the thermoelectric response of aqueous solutions.	


\newpage
\section*{1. Introduction}
Ludwig \cite{Ludwig} demonstrated in 1856 that thermal gradients can induce concentration gradients in aqueous solutions. Shortly after, Soret performed systematic investigations of aqueous solutions~\cite{Soret}, providing  a more complete picture of this effect. Following these two seminal works, many studies have quantified the Soret coefficient of salt solutions and aqueous suspensions. The investigation of the response of water to thermal gradients has also been considered more recently. It has been demonstrated that water gets polarized in the presence of thermal gradients~ \cite{bresme2008,armstrongbresme2015,Iriarte-Carretero2016}.

Experiments have shown that the Soret coefficient features a temperature inversion, with the coefficient changing sign at a specific temperature~\cite{alexander1954,Gaeta}. 
The temperature inversion signals a substantial modification in the response of the solution to a thermal gradient. The solution changes from thermophobic at high temperatures, to thermophilic at low temperature, with the salt accumulating in the hot region in the latter case. 
The existence of temperature inversion effects in alkali halide aqueous solutions has been confirmed using state of the art thermal diffusion force Rayleigh scattering (TDRFS)~\cite{blanco2011, kohler1995, Romer} techniques as well as computer simulations~\cite{Romer}. This phenomenology can be consistently reproduced both experimentally and theoretically, and it is widely accepted. In his seminal work, Brenner~\cite{brenner} proposed a connection betwee the sign change of the Soret coefficient and the thermal expansion of water. The latter changes sign at the maximum of density of water ($\sim 4^o$~C at 1 bar pressure), and it was proposed that this change in sign could be  correlated to the inversion effect. This idea is appealing, but it has not been fully supported by experiments~\cite{putnam2007} nor computer simulations~\cite{Romer}. Hence, a microscopic explanation is still sought.

A few experimental studies of aqueous solutions have reported the 
existence of a {\it minimum} in the Soret coefficient too. This is an interesting effect, since at the minimum the thermodiffusion response should be maximized. The minimum has been
observed in NaCl and KCl solutions~\cite{Gaeta}. 
For NaCl the minimum was observed at low concentrations $< 10^{-1}$~M, while for KCl it was found in the
range 0.1-1~M.  
In a recent study~\cite{Colombani}, a sharp minimum, at higher salt concentrations, closer to 1~M, was observed in LiCl solutions. In all these experiments the 
minima appears in the thermophilic regime, {\it i.e.}, when the Soret coefficient is negative, and the salt migrates preferentially to the hot region. 
Unlike the reversal in the sign of the Soret coefficient 
the minimum in the Soret coefficient has not been 
confirmed theoretically yet.

Early theories by Eastman and Agar~\cite{Eastman, Agar1989} introduced the concept of heat of transport and related this quantity to the Soret coefficient. These early works focused on the low dilution limit, making it difficult to extrapolate to finite concentrations, where non ideal effects in the Soret coefficient, such as minima and sign inversion, are observed. 
The heat of transport has been considered in modern studies of the Soret effect~ \cite{PhysRevLett.101.108302,PhysRevE.83.061403,PhysRevLett.108.118301}, and therefore its investigation and quantification remains of prime interest. In particular a good understanding of the dependence of the heat of transport on the thermodynamic conditions, temperature and salt
concentration can offer valuable insight to understand thermoelectric phenomena in aqueous solutions too. 
Indeed, the individual heats of transport of ions can be combined to estimate Seebeck coefficients~\cite{putnam}. This coefficient is important since it defines the strength of the thermolectric response of electrolyte solutions, and
it has been argued it may play a role in determining the thermophoretic response of colloidal suspensions~\cite{PhysRevLett.101.108302,PhysRevE.83.061403,PhysRevLett.108.118301,PhysRevLett.112.198101,putnam2007} and biomolecules~\cite{Duhr}.

In this work we aim to advance our ability to describe the heat of transport as a function of temperature and salt concentration. 
We have tackled this problem by using Non-Equilibrium Molecular Dynamics (NEMD) simulations. This approach has advanced significantly in the last years~\cite{bresme2015}, and it is now possible to quantify the Soret coefficients of aqueous solutions, and to reproduce their experimental response~\cite{nieto-draghi2005,Romer}. 
We have taken advantage of NEMD state of the art computational approaches 
to investigate the thermodiffusion of LiCl solutions. This solution  offers some interesting features that motivate our choice of system. Firstly, early studies indicated that at low concentrations the heats of transport of Li$^+$ and Cl$^-$ are equal. This means that in that regime the thermoelectric effects are irrelevant. However, at finite concentrations, \textit{i.e.} 10 mM, at 298~K the heat of transport of Li$^+$ and Cl$^-$ are different~\cite{TF9605601409} leading to not negligible thermoelectric effects~\cite{putnam}. Secondly, the thermodiffusion response of LiCl in the 
medium-high concentration regime ($>0.1$~M) is of great interest, since it was suggested recently that the Soret coefficient of LiCl features a minimum~\cite{Colombani}. 
These observations motivate us to apply our method to quantify the heat of transport of this salt. 
We will also establish correlations between the behavior of the ionic heats of transport and the thermoelectric response of this solution.

\section*{2. Materials and Methods}
\subsection*{2.1. Heat of transport and computational approach}

The Soret coefficient has been related to the heat of transport, $Q^*$. The works of Eastman and Agar are of particular significance in this instance~\cite{Eastman,Agar1989}. Eastman proposed that when a solute moves between regions at different temperatures, an amount of heat, $Q^*$, is absorbed or released 
in order to keep the temperature constant~\cite{Agar1989}. Eastman derived one equation connecting the heat of transport to the change of the chemical potential with concentration as well as with the gradient of concentration with temperature, {\it i.e.}, the Soret coefficient:

\begin{equation}
	\label{eq:qstar}
	Q^* = -\left(\frac{\partial \mu}{\partial b}\right)_{P,T} \left(\frac{db}{d \ln T}\right)_s
\end{equation}
\noindent
where the subscript $s$ refers to steady state conditions, and $\mu$ is the chemical potential of the solute and $b$ is the molality. 

This equation involves two quantities that can be accessed using computer simulations. The first term of the right hand side involves a chemical potential that we can compute using a perturbation approach under equilibrium conditions. The second term on the right hand side is connected to the Soret coefficient, and it can be computed using NEMD simulations.   

To calculate the heat of transport we performed NEMD simulations of LiCl aqueous solutions as a function of salt concentration and temperature. The simulations (NEMD) were conducted using the methodology discussed in reference~\citen{:/content/aip/journal/jcp/137/7/10.1063/1.4739855}. 
 
\begin{figure*}[ht]
\centering
\begin{center}
\begin{tabular}{lll}
\includegraphics[scale=0.23]{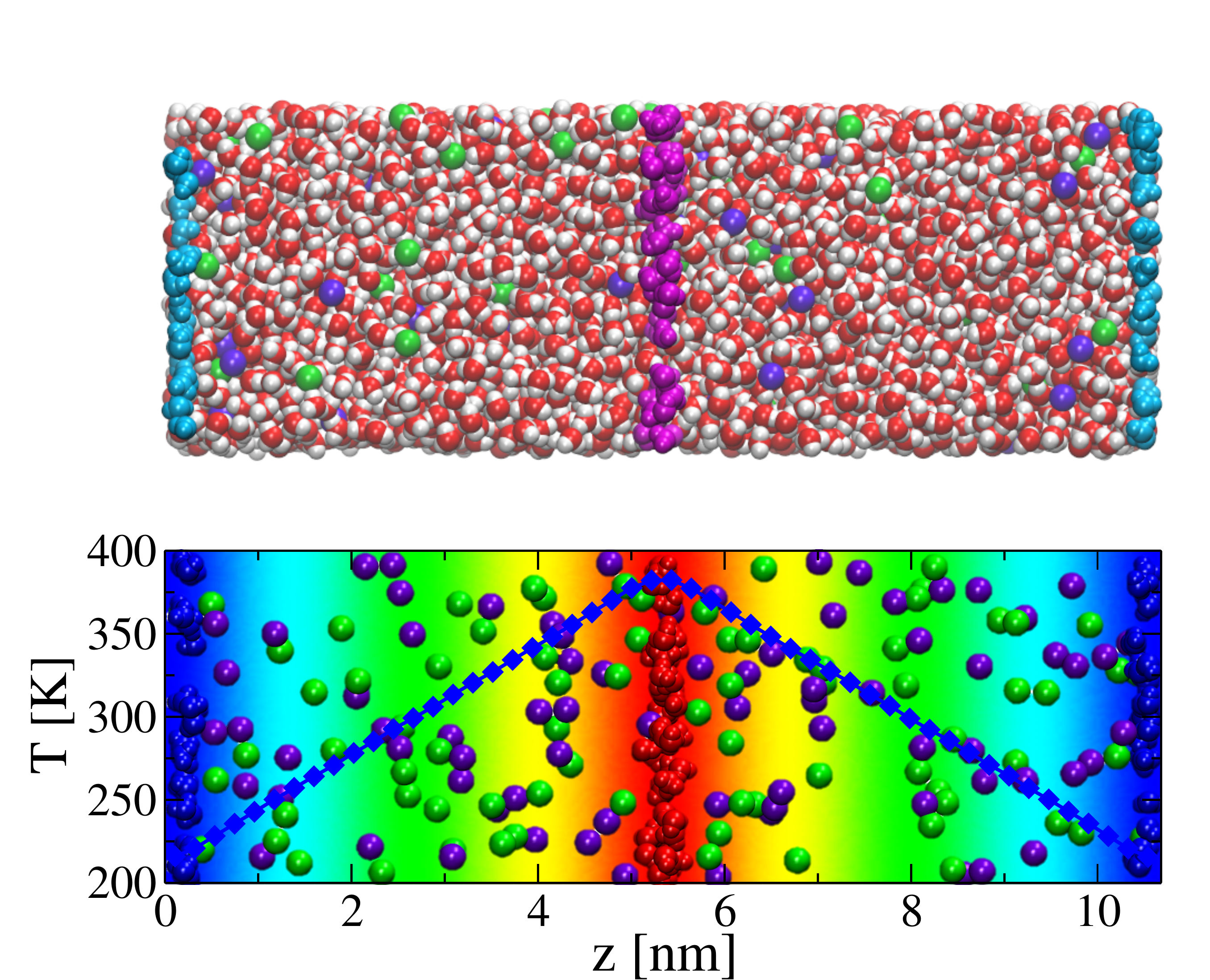} &
\includegraphics[scale=0.23]{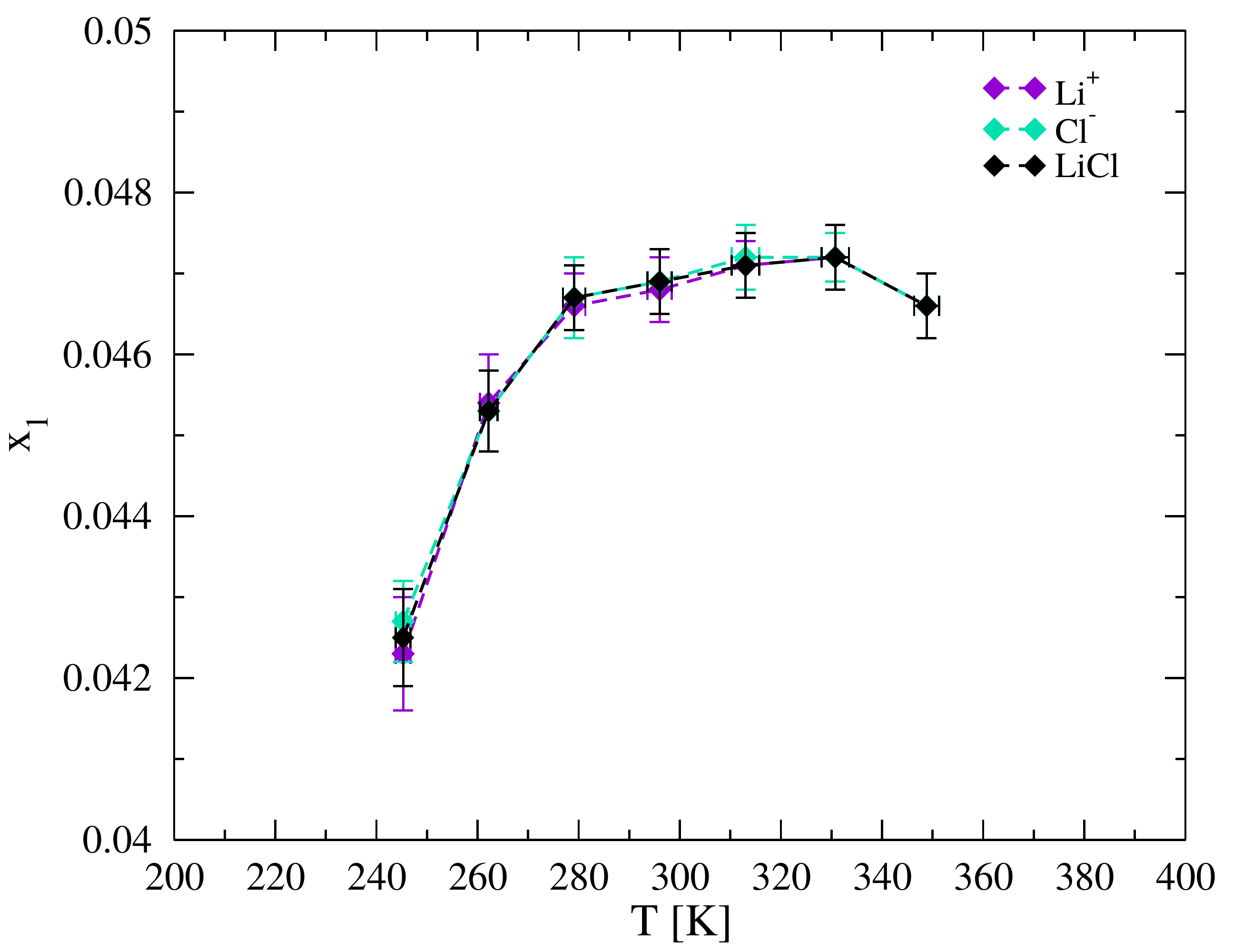} 
\end{tabular} 
\end{center}
\caption{(Top left) Snapshot of a LiCl solution under a thermal gradient showing the ions, (green -- Cl$^-$, violet -- Li$^+$) and water (white -- hydrogen, red -- oxygen). The thermostatting layers (see Methods section for more details) are highlighted in cyan (cold layer) and magenta (hot layer).  (Bottom left) and (right) panels show the  temperature of the unrestrained water molecules 
and the molar fraction dependence on the temperature,  at the stationary state,600 bar and average salt molality 2.5 mol kg$^{-1}$.}\label{fig:system}
\end{figure*}

In this method 
we define thermostatting regions (see Fig.~\ref{fig:system} - top left panel) where the temperature of the molecules is adjusted to predefined hot and cold values, while the rest of the molecules are not thermostatted, but adjust their temperature via interactions with the thermostatting molecules. For typical simulation cell sizes, this method readily produces a stationary heat flux in a few hundred picoseconds. 
The simulated system (see Fig.~\ref{fig:system}), consisted of a prismatic box with vectors, $\lbrace L_x, L_y, L_z \rbrace / L_x = \lbrace 1, 1, 3 \rbrace $, with $L_x =3.55$~nm. We used different number of water molecules, between 4306 and 4484, and a varying number of LiCl ion pairs between 77 and 385, in order to match the desired salt concentrations (1.0 -- 5.6~m).

The Li-Cl and LiCl-water interactions were calculated using a combination of Lennard-Jones and Coulombic interactions, and the cross interactions
were computed using standard combination rules. For water we employed the SPC/E model~\cite{Berendsen87} and for the ion-ion and ion-water interactions the model by Dang et al.~\cite{Dang1,Dang2,Dang3,Dang4}, which has been tested extensively in simulations of bulk and interfaces~\cite{Wynveen2010}. This model predicts a tetrahedral solvation structure for Li$^+$, which is compatible with predictions from accurate density functional theory computations~\cite{daub}. This coordination is close to that predicted
in neutron scattering experiments~\cite{doi:10.1021/jp511508n}. 
We performed simulations over 16~ns and the trajectories were analyzed to calculate composition and temperature profiles, which were later used to calculate Soret coefficients ($s_T$) from~\cite{Groot}, 

\begin{equation}
\label{eq:soret}
s_T = -\frac{1}{x_1 x_2} \left( \frac{\nabla x_1}{\nabla T} \right)_{J_1=0}   \approx  -\frac{1}{x_1}   \left( \frac{d x_1}{d T} \right)_{J_1=0},
\end{equation}
\noindent
where $x_1$ and $x_2$ represent the molar fraction of the salt and the solvent, and $J_i$ the mass flux of component $i$. 
Since the amount of solvent exceeds considerably the amount of salt it is convenient to use the approximation shown on the right hand side of equation~(\ref{eq:soret}), to calculate our $s_T$. 
The temperature dependence of our Soret coefficients was fitted to the empirical equation of Iacopini et al.~\cite{Iacopini06}. Further simulations details are provided in the Methods section. 

As shown in equation~(\ref{eq:qstar}), the computation of the heats of transport requires knowledge of the dependence of the chemical potentials of the ions as a function of salt concentration. We performed simulations in the NPT 
ensemble to obtain the excess chemical potential, $\mu_{ex}$ for the ions in the solvent~\cite{Ben-Naim}. This excess chemical potential corresponds to the work
required to transform the system from \textit{State 1} ($\mathcal{S}$1) containing $N_{i}$ ions and $N_{w}$ water molecules, to the \textit{State 2} ($\mathcal{S}$2) consisting of $N_{i}+1$ ions and $N_{w}$ water molecules. We used a perturbation method and a solute coupling parameter, $\lambda \in [0,1]$, which allows a smooth interpolation between $\mathcal{S}$1 and $\mathcal{S}$2. 
The excess chemical potential was then calculated using Kirkwood's formula~\cite{Kirkwood} in combination with the Bennett's acceptance ratio method~\cite{BENNETT1976245}. The chemical potential reported here quantifies the work required to move one ion from vacuum to the bulk solution, when this process is performed at constant pressure. We added to the excess chemical potential the corresponding ideal gas contribution $\mu_{id}$ of moving the ion from the gas phase at the selected pressure and volume
$V_1 = N_{ion} N_A k_B T / P$
at pressure, $P$, temperature, $T$ and average volume $\left<V\right>$~\cite{doi:10.1021/jp056043p}, to obtain the total chemical 
potential, 
\begin{equation}
\mu_{\alpha} = \mu_{ex, \alpha} + k_B T \ln \bigg ( \dfrac{N_{\alpha} k_B T}{P \left<V\right>} \bigg),
\label{eq:free_en_tot1}
\end{equation}
\noindent
where $N_{\alpha} = N_{LiCl} = N_{Li^+} = N_{Cl^-}$ is the total number of cation-anion pairs, 
and $k_B$ the Boltzmann constant. All our simulations were performed at 100 and 600 bar. We find little differences between the Soret coefficients obtained with these two pressures.

\subsection*{2.2. Computer simulation approach}

We performed all the NEMD and equilibrium simulations by using a modified version of GROMACS v.~4.6.3~\cite{gromacs}.

In the NEMD approach we restrained the position (in the direction of the heat flux -- $z$) of those oxygen atoms belonging to water molecules lying in the hot and cold thermostatting regions, at the beginning of the simulation. We used a harmonic potential with a force constant equal to 1000~ kJ~mol$^{-1}$~nm$^{-2}$. In our approach the restrained water molecules rotate freely, and they also translate in the $xy$ plane.  
The restrained molecules were thermostatted every time step using the  v-rescale algorithm~\cite{:/content/aip/journal/jcp126/1/10.1063/1.2408420}.

The cross interactions between different species were computed using standard combining rules: $\sigma_{\alpha \beta} = (\sigma_{\alpha \alpha}+\sigma_{\beta \beta})/2$, $\epsilon_{\alpha \beta} = \sqrt{\epsilon_{\alpha \alpha} \epsilon_{\beta \beta}}$. To model the water-water interactions we employed the SPC/E model~\cite{Berendsen87} while the model by Dang et al.~\cite{Dang1,Dang2,Dang3,Dang4} was chosen to compute the LiCl interactions. 
This model 
predicts water coordination numbers for Li$^+$ compatible with a tetrahedral arrangement of water molecules and is consistent with accurate density functional theory computations~\cite{daub}. Numerical values for the potential are given in Table \ref{tab:LJpar}. The Lennard-Jones interactions were truncated at  $r_c=1.5$~nm, and the Coulombic interactions were computed in full using the particle-mesh Ewald method (PME) with a mesh width of  $0.12$ nm and an interpolation order of $4$.   
The equations of motion were integrated 
with the leap-frog algorithm using a time step of 2~fs.  

A typical simulation involved a 1~ns pre-equilibration, in the NPT ensemble, of a box containing pure water at either $\sim$100~bar or $\sim$600~bar and  temperature, $T = (T_{COLD} + T_{HOT})/2$, where $T_{COLD}$ and $T_{HOT}$ are the temperatures in the NEMD simulations. The ions were then added to the desired concentration and the whole system was equilibrated again for 1~ns at the corresponding pressure and $T = (T_{COLD} + T_{HOT})/2$. 
 
Following the set up of the hot and cold regions of width $\simeq 0.1$ nm, the whole system was simulated by switching on the thermostats at temperatures $T_{COLD}$ and $T_{HOT}$ for several ns, to ensure the stationary state is reached. We then performed production runs of 16~ns. The trajectories were analyzed every 100 time steps to extract temperature, and concentration profiles, by dividing the simulation box in 100 sampling volumes 
along the direction of the thermal gradient, $z$. The temperature profile was calculated using the equipartition principle by sampling the velocities of the water molecules and the ions. The Soret coefficients were fitted to the empirical equation of Iacopini et al.~\cite{Iacopini06}, $s_T(T) = s_{T}^{\infty} \left[1 - \exp\left( (T_{0} - T)/\tau \right)\right]$, which describes accurately the temperature dependence of $s_T$. $s_T^{\infty}$ and $T_0$ represent the asymptotic limit of $s_T$ and the inversion temperature, respectively, and $\tau$ is a parameter that determines the temperature dependence of the Soret coefficient. Using equation~(\ref{eq:soret}) along with the equation for $s_T(T)$ above,~\cite{Iacopini06} we derived an equation for the temperature dependence of the concentration, 

\begin{equation}
\label{eq:ct}
b(T) = b_{0} \exp{[ -s_{T}^{\infty}(T + \tau \mathrm{e}^{\frac{T_{0} - T}{\tau}} + k)]}
\end{equation}
where $b(T)$ is the molality at temperature $T$, $b_0$ is the average molality of the solution and $s_{T}^{\infty}$, $T_{0}$, $\tau$ and $k$ are fitting parameters. Equation~(\ref{eq:ct}) provides an excellent fitting of all our NEMD simulation data. The standard deviation of the concentration profiles and $s_T$ at $\sim$100 and 600~bar were obtained from the analysis of 20 independent trajectories (16 ns each).

\begin{table}[h!]
\begin{center}
\begin{tabular}{|  c  |  c  |  c  |  c  |  c  |  c  |  c  |}
\hline
 atom type & $mass \; [au]$ & $\sigma \; \bigl[ \stackrel{\circ}{A} \bigr]$  &  $\varepsilon \; [kJ/mol]$ & $q(e)$ \\
\hline
$Li^{+}$ & 6.941 &  1.506 & 0.6904 & +1.0000\\
$Cl^{-}$ & 35.453 & 4.401 & 0.4184 & -1.0000\\
$OW$ & 15.9994 & 0.3166 & 0.65 & -0.8476\\
$HW$ & 1.0 & 0.0 & 0.0 & +0.4238\\
\hline
\end{tabular}
\end{center}
\caption{Lennard-Jones parameters describing the interactions between the ions Li$^+$, Cl$^{-}$ and the water molecule. The parameters for the ions are taken from Dang et al.~\cite{Dang1, Dang2, Dang3, Dang4}, and for water from the SPC/E model~\cite{Berendsen87}.}\label{tab:LJpar}
\end{table}

\subsection*{2.3. Free energy computations}

In our computations we decoupled the van der Waals and the Coulombic contributions 
by considering a thermodynamic path where first a neutral Lennard-Jones atom in the solution is created, which is then fully charged, $q = \pm 1 e$. 

Previous works dealing with chemical potential computations have discussed corrections to the free energy of solvation of the ions, which need to be included in non-neutral systems when the computations are performed using the Ewald Summation method~\cite{Hummer, Netz}. The correction factor decreases as the simulated box size increases, since it is connected to the electrostatic interactions between periodic images. 
Unlike in many previous studies concerned with ionic solvation free energies, our computations are performed in an aqueous solution at finite concentrations. We expect that the additional salt will screen the electrostatic interactions. Computations of NaCl at typical concentrations studied here, $>1$~M, and 300 K, revealed a weak size dependence of the electrostatic contribution to the chemical potential of Na$^+$. For box sizes 3, 4 and 6 nm we found, -370.2 $\pm$ 0.1, -370.1 $\pm$ 0.1 and -370.0 $\pm$ 0.1 kJ~mol$^{-1}$, which are within the uncertainty of our computations.  This observation agrees with previous studies of KF aqueous solutions~\cite{Ferrario}, who did not include corrections to the chemical potential. Therefore, we did not include additional corrections in our chemical potential results.

The chemical potential reported in our work quantifies the work required to move one ion from vacuum to the bulk solution at constant pressure. We added to the excess chemical potential the corresponding ideal gas contribution $\mu_{id}$ 
at pressure, $P$, temperature, $T$ and average volume $\left<V\right>$:\cite{doi:10.1021/jp056043p}

\begin{equation}
\mu_{\alpha} = \mu_{ex, \alpha} + k_B T \ln \bigg ( \dfrac{N_{\alpha} k_B T}{P \left<V\right>} \bigg)
\label{eq:free_en_tot}
\end{equation}
\noindent
where $N_{\alpha} = N_{LiCl} = N_{Li^+} = N_{Cl^-}$ is the total number of cation-anion pairs, 
and $k_B$ is the Boltzmann constant.
In order to use the perturbation approach and calculate the chemical potential of the cation, $\mu_{Li^+}$ and the anion $\mu_{Cl^-}$ we used 30 $\lambda_{vdw}$ values for the growth of the Lennard-Jones spheres and 20  $\lambda_c$ values for the charging process of Li$^+$ and Cl$^-$, respectively. The chemical potential computation were performed in LiCl aqueous solutions at the desired concentrations. 
The chemical potentials for Li$^+$ and Cl$^-$ ions were calculated separately by using two independent simulations sets. The total chemical potential  
$\mu_{LiCl}= \mu_{Li^+} + \mu_{Cl^-}$ was then computed by adding the anion and cation contributions.
For each $\lambda$, the simulations were performed in the NPT ensemble using a time step of 2~fs. A typical simulations involved a 5~ns equilibration period, followed by a 40~ns to 80~ns production period. We discarded the first 1~ns of the trajectories. 
We used the v-rescale thermostat with a time constant of 0.1~ps and the Parrinello-Rahman barostat, with time constant 1~ps. In the chemical potential computations, the cutoff radius for the Lennard Jones and the Coulombic potentials and for the neighbor list were set to 0.9~nm and the neighbor list was updated every step. 
We tested the impact of the cutoff on the chemical potentials. We find that simulations with 0.9 and 1.5~nm cutoffs predict chemical potentials within the statistical uncertainty of the computations, since the
chemical potentials are dominated by the electrostatic contribution, which is treated in full.
The system sizes used for each concentration  used to compute the chemical potential are reported in the Supporting Information. We tested that our simulation set up produced chemical potentials consistent with those published by different authors for NaCl salts using the same ion and water force fields~\cite{Mester2015}.

To calculate the thermodynamic factor using the equation~(\ref{eq:gammachempot}), we fitted our chemical potentials to the equation 
~\cite{Mester2015},

\begin{equation}
	\label{eq:fitting}
\mu = K + k_B T N_A \left[ 2 \ln b + 2 \ln_{10} \left(\frac{-A \sqrt{b}}{1 + B\sqrt{b}} + \beta b + Cb^2 + Db^3 \right) \right] 
\end{equation}
where $K, A, B, \beta, C, D$ are constants and $N_A$ is the Avogadro's number. The fitting parameters are reported in the Supporting Information. 
Our values of the chemical potential for Cl$^-$ are noisier than those of Li$^+$. To improve the quality of the fitting curve we 
fitted first $\mu_{LiCl}(b)$ and $\mu_{Li^+}(b)$ and then extracted the fitted function for the chemical 
potential for the anion from $\mu_{Cl^-}(b) = \mu_{LiCl}(b) - \mu_{Li^+}(b)$, in this way we overcome the 
appearance of unphysical oscillations that can result from the direct fitting using equation~(\ref{eq:fitting}). The resulting
fitting 
interpolates well our simulation data.

\section*{3. Results}
\begin{figure}[ht!]
\begin{center}
\begin{tabular}{cc}
\includegraphics[scale=0.24,natwidth=610,natheight=642]{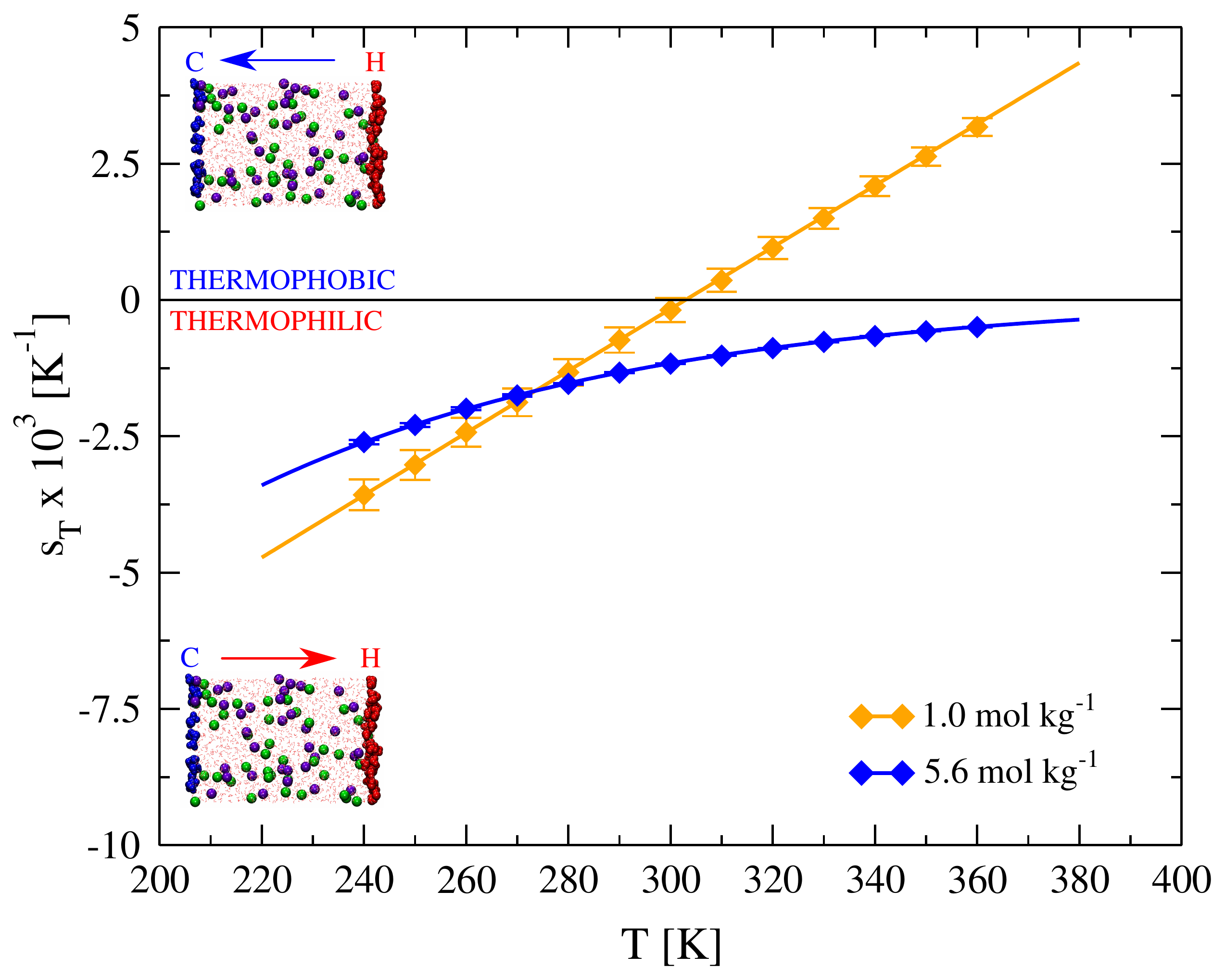} &
\includegraphics[scale=0.24,natwidth=610,natheight=642]{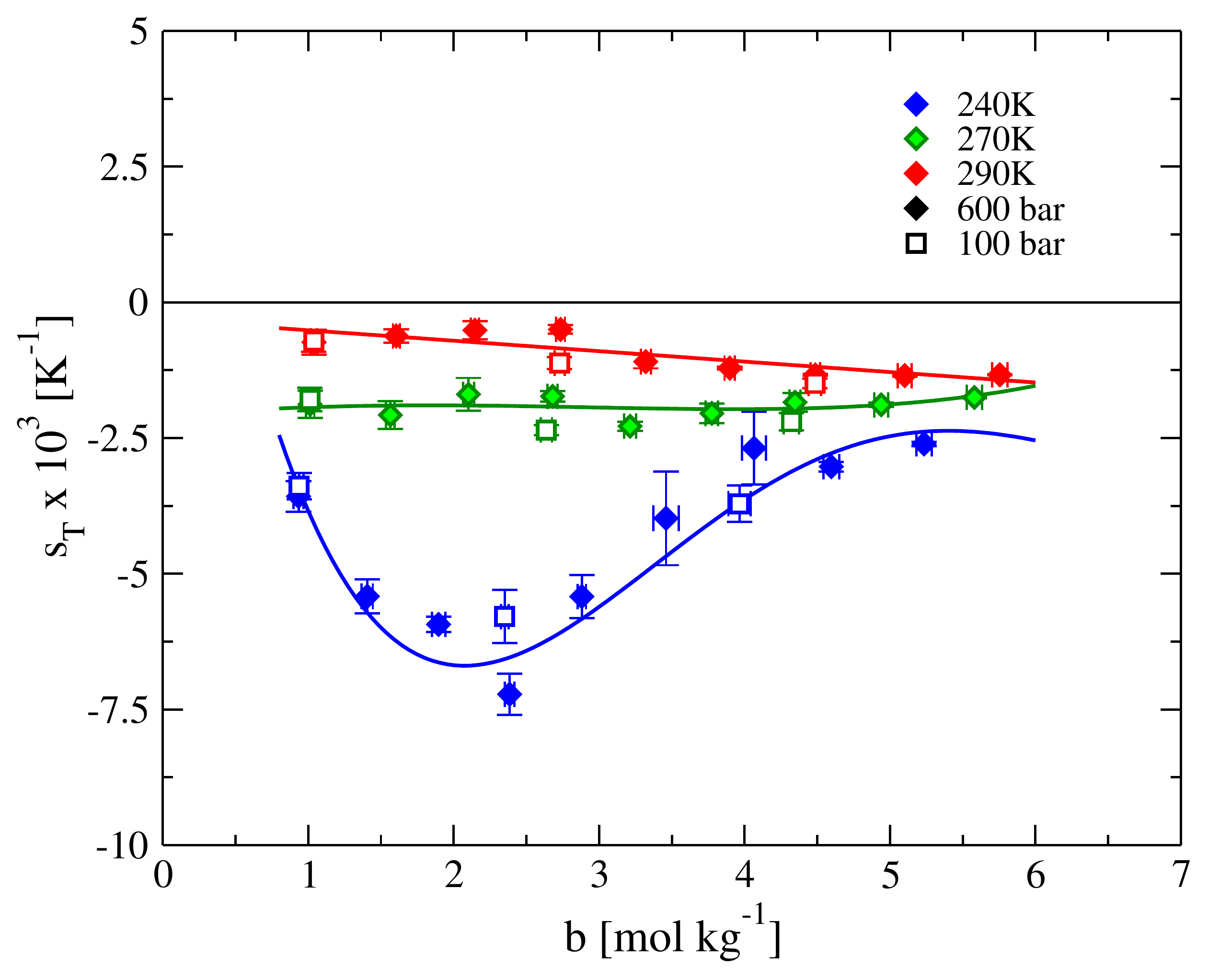}
\end{tabular}
\end{center}
\caption{(Left) Temperature dependence of the Soret coefficient as a function of the LiCl salt concentration at 600~bar. The diamonds and full lines represent our NEMD data, while the triangles and dashed lines represent the experimental Soret coefficients reported in reference~\citen{Colombani}. (Right) Soret coefficient of LiCl solutions as a function of molality along different isotherms, as specified in the legend. The full diamonds and empty squares represent the NEMD simulations performed at 600 and 100~bar, respectively. The lines are just a guide to the eye. \label{fig:conc2}}
\end{figure}

When the aqueous solution reaches the stationary state, a constant heat flux, temperature, density and concentration gradients are established. We show in Fig.~\ref{fig:system} 
representative results for the temperature along the simulation box and the solute mole fraction in the temperature range considered. We computed the Soret coefficient from the analysis of the temperature and the molar fraction profiles (see Fig.~\ref{fig:system} and the Methods section for further details), using equation~(\ref{eq:soret}). We show in Fig.~\ref{fig:conc2} the dependence of the simulated $s_T$ for concentrations 1~m and 5.6~m as a function of temperature. The magnitude of our $s_T$, $\sim 10^{-3}$~K$^{-1}$ is in the range of the available experimental data for
alkali halide solutions~\cite{Gaeta,Romer} and for LiCl solution~\cite{Colombani} at similar concentration. 
Our results at 1~m feature the inversion effect, namely, the Soret coefficient changes sign at a specific temperature. At that temperature the response of the solution to the thermal gradient changes from thermophilic (at low temperature) to thermophobic (at high temperature). 
The strong dependence of $s_T$ with salt concentration reported here agrees with previous experimental observations~\cite{Colombani} (see Fig.~\ref{fig:conc2} - left). At the high concentrations, ~5.6 m, the aqueous solution is thermophilic ($s_T < 0$) in the whole temperature interval. This behavior is again consistent with the experimental observations, which also reported an overall thermophilic response for LiCl solutions at high salt concentrations~\cite{Colombani}.   

We tackle in the following the computation of the heat of transport, $Q^*$. $Q^*$ is a complex property that is defined by the interplay of electrostatic interactions, local energy changes associated with the interactions between the moving ion and the solute, and the breaking - reconstructing effect of the ion on the solvation water~\cite{Gaeta}. We have used our Soret coefficients, $s_T$, and chemical potential data, $\mu_{\alpha}$,  to quantify the individual ionic 
contributions to the heat of transport $Q^{*}_{\alpha}$, for $\alpha = (Li^+, Cl^-)$. To connect the total and ionic heats of transport of LiCl we use 
the fact that this is an additive property of the ions~\cite{Agar1960}, at least, for solutions of 1:1 electrolytes such as the ones investigated here, hence 
$Q^{*} = Q^{*}_{Li^+} + Q^{*}_{Cl^-}$.

Equation~(\ref{eq:qstar}) can be rewritten in terms of the Soret coefficient, $s_T$, and the thermodynamic factor, $\Gamma$, 
\begin{equation}
	\label{eq:qstar2}
Q^* = 2 s_T\ R\ T^2 \Gamma	
\end{equation}
\noindent
where
\begin{equation}\label{eq:gamma}
\Gamma =  1 + \left(\frac{\partial \ln \gamma_{+-}}{\partial \ln b} \right)_{P,T}  
\end{equation}

\noindent
where $\gamma_{+-}$ is the mean activity coefficient and $R$ the gas constant.
The thermodynamic factor is related to the chemical potential by,

\begin{equation}
	\label{eq:gammachempot}
	\Gamma = \frac{b}{2 R T} \left( \frac{\partial \mu}{\partial b} \right)_{P,T}
\end{equation}
\noindent
where $\mu = \mu_{Li^+} + \mu_{Cl^-}$. The individual chemical potentials can also be used to obtain the individual ionic heats of transport,

\begin{equation}
	\label{eq:ionicheat}
	Q^*_{\alpha} = T\ b\ s_{T,\alpha} \left(\frac{\partial \mu_{\alpha}}{\partial b} \right)_{P,T}
\end{equation}

Our simulations show Soret coefficients for the cations and anions that are indistinguishable within the uncertainty of the computations, hence we used $s_{T,Li^+}  = s_{T, Cl^-}  = s_T $. This point cannot be tested in experiments, since there are not experimental 
approaches that provide  measurements of the Soret coefficients for individual ions. Hence the simulations provide additional insight into the behavior of this coefficient. To obtain the heat of transport we fitted the chemical potentials to a polynomial function as explained in the Methods section. This derivative was subsequently employed in equations (\ref{eq:qstar2}) and (\ref{eq:ionicheat}). The chemical potentials (see Supporting Information for numerical data) are consistent  
with estimated data of alkali halide ions in water at infinite dilution, which are typically of the order of $\sim -371~$kJ~mol$^{-1}$ for chloride in water~\cite{doi:10.1021/jp951011v} and $\sim -480$~kJ~mol$^{-1}$ for lithium in water~\cite{doi:10.1021/jp056043p} at 298~K. 

\begin{figure}[ht!]
\centering
\includegraphics[scale=0.3,natwidth=610,natheight=642]{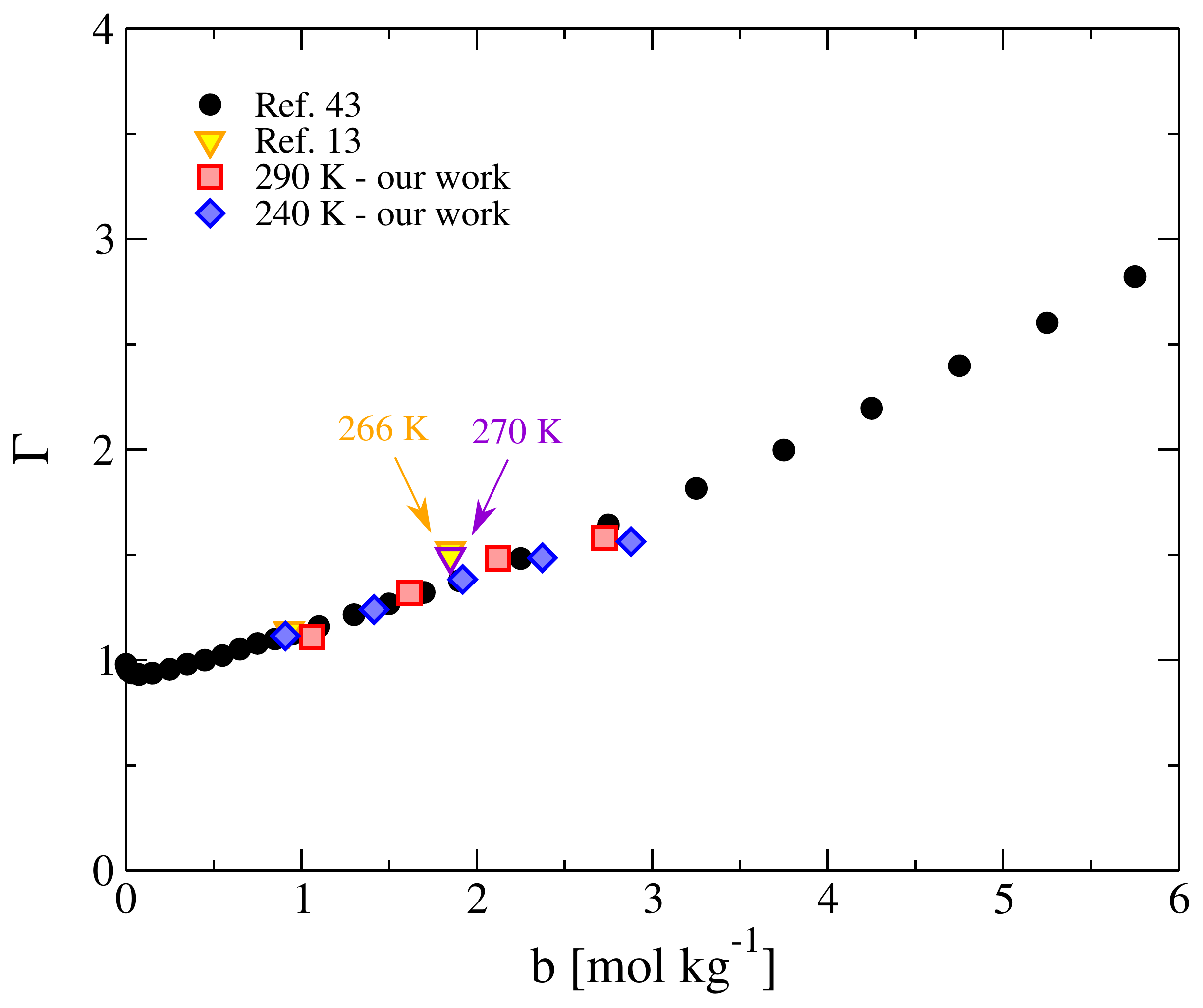}
	\caption{Thermodynamic factor as a function of the molality. The filled circles represent the experimental data from Harmer et al.~\cite{:/content/aip/journal/jpcrd/1/4/10.1063/1.3253108} at 300 K, while the yellow square are the data extrapolated from Colombani et al.~\cite{Colombani}; the diamonds and squares represent the data from this work at 240~K (blue) and 290~K (red), respectively.}\label{fig:Gamma}
\end{figure}

We compare in Fig.~\ref{fig:Gamma} the thermodynamic factor at 290~K, calculated from our chemical potentials, using the equation~(\ref{eq:gamma}), and the experimental data from Harmer et al.~\cite{:/content/aip/journal/jpcrd/1/4/10.1063/1.3253108}. Our thermodynamic factors feature the typical increase with ion concentration and are of the same order as typical experimental data for alkali halides at similar temperatures~\cite{:/content/aip/journal/jpcrd/1/4/10.1063/1.3253108}.

\begin{figure}[h!]
\begin{center}
\begin{tabular}{cc}
 \includegraphics[scale=0.35]{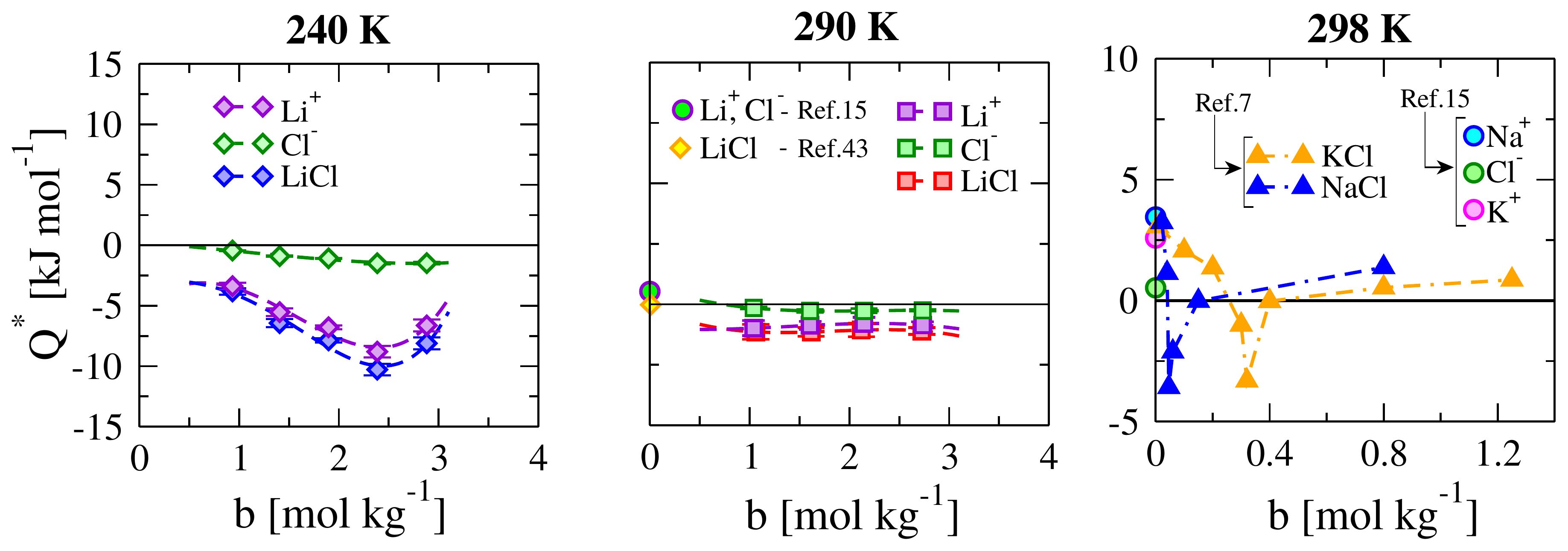} &  
\end{tabular}
\end{center}
\caption{Heat of transport as a function of the salt concentration and temperature for different salt solutions, as specified in the legend. Data for $Q^*$, $Q^*_{Li^+}$ and $Q^*_{Cl^-}$ are shown, for 240~K (left) and 290~K (middle), respectively. The yellow diamond in the middle panel represent experimental data from reference~\citen{:/content/chem/24/1/AJA03794350_1901} at $\sim$298~K. The circle in the middle and in the right panels represent data at infinite dilution and 298~K from reference~\citen{Agar1989}. The panel on the right shows experimental estimates of $Q^*_{NaCl}$ and $Q^*_{KCl}$ at 298~K from reference~\citen{Gaeta}.}
\label{fig:qstar}
\end{figure}

We have computed the individual heats of transport $Q^*_{\alpha}$ for $\alpha = (Li^+, Cl^-)$, from the derivative, $(d\mu_{\alpha}/db)_{PT}$, using as input our simulated chemical potentials, the simulated Soret coefficients and equation~(\ref{eq:qstar}) to model the concentration dependence of $\mu_{\alpha}$. As noted above the same $s_T$ was used for cations and anions. The derivative of the chemical potential, $d\mu_{\alpha}/db$, were calculated using the fitting curves obtained through equation~(\ref{eq:fitting}). Fig.~\ref{fig:qstar} shows our
heats of transport at 290~K. It is instructive to compare our predictions with values reported in the literature. The estimates of $Q^*$ by Agar et al.~\cite{Agar1989} at 298 K, 1 bar (calculated using the Born theory) are of the order of 0.53~kJ mol$^{-1}$ for Li$^+$ and Cl$^-$. Our heats of transport at 290~K are similar in magnitude, $\sim$ 1.14 - 1.10 kJ mol$^{-1}$ in absolute value for $\sim$1-3~m, but have opposite sign. Our simulations show that unlike in the infinite dilution estimate, the heats of transport of anions and cations are different, with that of Li$^+$ being about 3-4 larger than that of Cl$^-$. 
In another study Gaeta et al.~\cite{Gaeta} estimated values for the heats of transport obtaining Q$^* \sim 1.38$ kJ~mol$^{-1}$ for 0.80~M NaCl and $\sim 0.86$ kJ~mol$^{-1}$ for 1.25~M KCl aqueous solutions at 303.15 K, again of the same order as the Q$^*$ values found here. The estimates of heats of transport for NaCl and KCl at 1~M concentration reported by Gaeta et al.~\cite{Gaeta} were also different from the estimates at infinite dilution using Agar's approach. This confirms that salt concentration influences the heats of transport, a conclusion consistent with our simulations. 

Colombani et al.\cite{Colombani} investigated the Soret coefficient of LiCl in a range of concentrations and temperatures similar to the one investigated here, although they did not report heats of transport. We have therefore analyzed the existing experimental data and estimated Q$^*$.  
Extrapolating the experimental $s_T$ to 273 K, and the thermodynamic factor $\Gamma$ values from reference~\citen{:/content/aip/journal/jpcrd/1/4/10.1063/1.3253108} we estimate that $Q^*$ of LiCl varies between -6.58~kJ~mol$^{-1}$ for 1.85 m concentration and -12.57~kJ~mol$^{-1}$ for 0.56 m. These values are much higher than the ones reported for NaCl or KCl, and they are negative, therefore in agreement with the sign of the heats of transport calculated with our method. Considering together the experimental estimates and our data we conclude that the heat of transport can vary significantly in sign and magnitude with respect to the values estimated at infinite dilution. We cannot compare the individual heats of transport of ions with experiment, since these cannot be extracted from experimental studies. In fact, the possibility of estimating individual heats of transport is a major strength of our
approach. 
We have used our individual heats of transport to estimate the Seebeck coefficient, $S_e$, of the solutions. To obtain these coefficients we used the relationship $S_e = (Q^*_+ - Q^*_-)/(2\ T\ e) $ (see {\it e.g.} reference~\cite{Groot}), where $e$ is the electron charge. This equation assumes that the transference number~\cite{Groot} of each component, which depend on the ionic mobility, is $1/2$ for both ions.  
The simulated Seebeck coefficients (see Table~\ref{tab:qstar290}) are in the range expected for this quantity, namely,  $\sim k_B/e$.  

At the lower concentration investigated, 1.0~m, and T = 290~K, we get, $S_e = -14.96 \pm 5.62~\mu V~K^{-1}$. This value is of the order of the Seebeck coefficient of NaCl using the recommended data by Agar et al.~\cite{Agar1989} at infinite dilution, but it is different from the Seebeck coefficient for LiCl at infinite dilution~\cite{Agar1989}, which would be zero according to Agar estimates. Our work again reveals substantial differences between the transport coefficients at infinite dilution and finite concentrations, which must be connected to the ionic correlations and deviations from ideality (see Fig.~\ref{fig:Gamma}). A temperature dependence of the Seebeck coefficient 
is also expected based on previous investigations~\cite{C3SM52779D}. 

\begin{table}[h!]
	\begin{center}
		\begin{tabular}{| c  |  c  |  c  |  c  |  c  |  c  |  c  |}
			\hline
			\ $T $ & $ b$ & $Q^*_{Li^+} $ & $Q^*_{Cl^-} $ &  $Q^* $ & $S_{e}$\\
			\hline
			240	&	$	0.93	$	&	$	-3.38	\pm	0.27	$	&	$	-0.44	\pm	0.03	$	&	$	-3.82	\pm	0.27	$	&	$	-63.54	\pm	5.83	$	\\
	&	$	1.41	$	&	$	-5.53	\pm	0.32	$	&	$	-0.90	\pm	0.05	$	&	$	-6.43	\pm	0.32	$	&	$	-99.95	\pm	7.02	$	\\
	&	$	1.90	$	&	$	-6.78	\pm	0.16	$	&	$	-1.08	\pm	0.03	$	&	$	-7.86	\pm	0.16	$	&	$	-123.05	\pm	3.50	$	\\
	&	$	2.39	$	&	$	-8.80	\pm	0.47	$	&	$	-1.49	\pm	0.08	$	&	$	-10.28	\pm	0.47	$	&	$	-157.85	\pm	10.19	$	\\
	&	$	2.88	$	&	$	-6.63	\pm	0.49	$	&	$	-1.48	\pm	0.11	$	&	$	-8.11	\pm	0.50	$	&	$	-111.24	\pm	10.80	$	\\

\hline

290	&	$	1.06	$	&	$	-0.99	\pm	0.31	$	&	$	-0.15	\pm	0.05	$	&	$	-1.14	\pm	0.31	$	&	$	-14.96	\pm	5.62	$	\\
	&	$	1.62	$	&	$	-0.88	\pm	0.17	$	&	$	-0.26	\pm	0.05	$	&	$	-1.14	\pm	0.18	$	&	$	-11.11	\pm	3.24	$	\\
	&	$	2.12	$	&	$	-0.81	\pm	0.26	$	&	$	-0.26	\pm	0.08	$	&	$	-1.07	\pm	0.28	$	&	$	-9.79	\pm	4.94	$	\\
	&	$	2.73	$	&	$	-0.86	\pm	0.14	$	&	$	-0.25	\pm	0.04	$	&	$	-1.10	\pm	0.14	$	&	$	-10.90	\pm	2.52	$	\\

\hline
		\end{tabular}
	\end{center}
	\caption{Heats of transport (in kJ mol$^{-1}$) and Seebeck coefficients (in $\mu V K^{-1}$) as a function of the salt molality (in mol kg$^{-1}$) and temperature (in K).}\label{tab:qstar290}
\end{table}

One interesting aspect of thermodiffusion behavior of alkali halide solutions is the experimental observation of minima in the Soret coefficient~\cite{Gaeta,Colombani}. We do not find evidence for a minimum at 290 K using our simulation model (see Figure~2). We therefore performed additional simulation scanning different temperatures. Fig.~\ref{fig:conc2} shows two additional isotherms, which show distinctive changes in the behavior of the Soret coefficient with temperature. At 270~K our model predicts a Soret coefficient that is essentially independent on temperature, and 240~K (25~K higher than the melting temperature of SPC/E water at 1 bar pressure~\cite{vegasanz}) we find evidence for a clear minimum. 
To the best of our knowledge this is the first time that a minimum in the Soret coefficient is observed using a theoretical approach. This is an important result of our work, since the observation of minima in experiments is restricted to a few experiments. A minimum in the Soret coefficient implies a maximization of the thermodiffusive response of the solution.
 We find the pressure does not influence significantly the Soret coefficient in the interval 100 to 600 bars (see Fig.~\ref{fig:conc2} - right panel). 

How do the heat of transport and the Seebeck coefficient change near the minimum of the Soret coefficient? We have tackled this question using our computational approach. Following the analysis of the high temperature system, we computed the thermodynamic factor at 240~K (see Fig.~\ref{fig:qstar}). We find that this
quantity does not depend significantly with temperature. Such behavior is in reasonable agreement with the one estimated on can infer from existing experiments of aqueous solutions (see Fig.~3). 

Our results for the heats of transport are reported in Table~\ref{tab:qstar290} and Fig.~\ref{fig:qstar}. Unlike in the high temperature case (c.f. results for 290 K and 240 K in Fig.~\ref{fig:qstar}), the heat of transport is found to change significantly near the minimum. Again we find large differences between the Li$^+$ and Cl$^-$ heats of transport, with the Li$^+$ contribution being much stronger than the Cl$^-$ one. These results highlight again the differences between the heats of
transport at finite concentrations and infinite dilution. The increase in the magnitude of the heat of transport near the minimum of the Soret coefficient is consistent with experimental analyses of NaCl and KCl solutions (see Fig.~\ref{fig:qstar}). Using our approach, we
can go one step further and obtain the individual contributions of the ions to the heat of transport. Our calculations show that Li$^+$
contributes significantly more to the heat of transport than Cl$^-$. The magnitude of the heats of transport of LiCl is larger than the one estimated experimentally for NaCl and KCl. Again we did not find experimental data for the heat of transport of LiCl. Hence, we reanalyzed  the existing data for Soret coefficients and thermodynamic factors and estimated the heat of transport of LiCl near the minimum of $s_T$. 
We find that the experimental heats of transport for LiCl at the minimum should be stronger than those for NaCl or KCl. The estimated data show good agreement with our simulation predictions. 
$Q^*$ has been identified before with the entropy transported by the solute, $S^*$~\cite{Eastman,Agar1989,WURGER2013438}. Considering this connection our results imply that the transported entropy of LiCl is higher that for other salts. Further we show that the main contribution to the observed behavior is associated to the transport of the Lithium cation.

Finally, we used the individual ionic heats of transport to examine the dependence of the Seebeck coefficient with salt concentration (see Table~\ref{tab:qstar290}). We find that the coefficient increases significantly near the minimum of the $s_T$. Our results indicate that the thermoelectric response of LiCl might be maximized at thermodynamic conditions corresponding to the minimum of the Soret coefficient. We estimate a Seebeck coefficient of $-157.85 \pm 10.19~\mu V K^{-1}$, which is definitely larger than the result obtained at  
higher temperatures. 

\section*{3. Conclusions}

We have proposed a computational approach to compute the heats of transport of aqueous solutions. Our approach is suitable to investigate solutions at intermediate concentrations, where non-ideal effects become important, and for
which there are no predictive theoretical approaches. Our method relies on the computation of Soret coefficients using Non-Equilibrium Molecular Dynamics simulations and chemical potentials using equilibrium simulations.  

We have applied this method to investigate the heats of transport of LiCl solutions as a function of interaction strength and temperature. The Soret coefficients computed in our work are consistent with existing experimental data of alkali halide solutions, in terms of magnitude and thermophilic character. They also reproduce the inversion behavior observed experimentally in a variety of systems, where the solution change from thermophilic to thermphobic at a specific temperature. We have reported the first theoretical demonstration of a minima in the Soret coefficient of aqueous solutions. This result supports the existence of such physical behavior in the Soret coefficient, which was reported so far in a very limited number of experiments. 

Although further analyses of specific systems at quantitative level may require more
involved forcefields, we find the following key conclusions, which should be taken into consideration for future experimental and theoretical studies:

$\bullet$ The heat of transport, $Q^*$, at finite concentrations is found to depend both on concentration and temperature. $Q^*$ can be significantly different from the heats of transport that have been estimated theoretically at infinite dilution, and that have been used to interpret thermodiffusion at finite concentrations.
The heat of transport of LiCl can be of opposite sign and larger than the estimates in the zero concentration limit. A re-analysis of existing experimental data, allowed us to validate our simulation predictions. 

$\bullet$ Our approach provides a theoretical route to calculate heats of transport of individual ions as a function of concentration and temperature. At finite concentrations the heat of transport of Li$^+$ differs substantially from that of Cl$^-$. This result deviates from the expected values at infinite dilution, where the heats of transport of Li$^+$ and Cl$^-$ are identical. Such deviations should be connected to the increasing role of non ideal effects, as demonstrated by the thermodynamic factor. Our computations show that Li$^+$ provides the larger contribution to the heat of transport of the solution. 

$\bullet$ The calculation of individual heats of transport of ions allows the estimation of the Seebeck coefficients, and therefore a quantification of thermoelectric effects in solutions. The Seebeck coefficient is found to change significantly with temperature and salt concentration. The thermoelectric response is maximized at thermodynamic conditions corresponding to the minimum of the Soret coefficient, where the Seebeck coefficient reaches values of the order of 100 $\mu$V/K in absolute value. The minimum of the Soret
effect is therefore a relevant physical phenomenon that may influence significantly the thermolectric behavior of solutions. More experimental work focusing on thermoelectric phenomena at experimental conditions compatible with minima in the Soret coefficient would be desirable.



\section*{Supporting Information}

\renewcommand{\thetable}{S\arabic{table}}
\renewcommand{\arraystretch}{1.4}
\setcounter{table}{0}

\begin{table}[ht]
\centering
		\begin{tabular}{ |  c  |  c  |  c  |  c  |  c  | }
			\hline
			 $T \, [K]$ & $ b\; [ kg \;mol^{-1} ]$ & $S_T \times 10^3 \; [K^{-1}]$ \\
			\hline
	240		&	$	0.934	\pm	0.034	$	&	$	-3.57	\pm	0.28	$	\\
			&	$	1.406	\pm	0.038	$	&	$	-5.42	\pm	0.31	$	\\
			&	$	1.896	\pm	0.046	$	&	$	-5.93	\pm	0.14	$	\\
			&	$	2.385	\pm	0.032	$	&	$	-7.22	\pm	0.38	$	\\
			&	$	2.881	\pm	0.029	$	&	$	-5.42	\pm	0.40	$	\\
			&	$	3.46	0	\pm	0.086	$	&	$	-3.98	\pm	0.86	$	\\
			&	$	4.065	\pm	0.082	$	&	$	-2.69	\pm	0.67	$	\\
			&	$	4.596	\pm	0.054	$	&	$	-3.03	\pm	0.09	$	\\
			&	$	5.234	\pm	0.054	$	&	$	-2.61	\pm	0.04	$	\\
			\hline
			270		&	$	1.013	\pm	0.029	$	&	$	-1.878	\pm	0.25	$	\\
			&	$	1.565	\pm	0.029	$	&	$	-2.08	\pm	0.26	$	\\
			&	$	2.102	\pm	0.038	$	&	$	-1.69	\pm	0.30	$	\\
			&	$	2.68	0	\pm	0.034	$	&	$	-1.73	\pm	0.10	$	\\
			&	$	3.213	\pm	0.042	$	&	$	-2.28	\pm	0.08	$	\\
			&	$	3.775	\pm	0.042	$	&	$	-2.05	\pm	0.18	$	\\
			&	$	4.347	\pm	0.038	$	&	$	-1.84	\pm	0.17	$	\\
			&	$	4.939	\pm	0.049	$	&	$	-1.89	\pm	0.04	$	\\
			&	$	5.58	0	\pm	0.053	$	&	$	-1.75	\pm	0.02	$	\\
			\hline		
	290 &	$	1.04	0	\pm	0.025	$	&	$	-0.74	\pm	0.23	$	\\
			&	$	1.606	\pm	0.023	$	&	$	-0.62	\pm	0.12	$	\\
			&	$	2.146	\pm	0.028	$	&	$	-0.52	\pm	0.17	$	\\
			&	$	2.735	\pm	0.03	0	$	&	$	-0.50	\pm	0.08	$	\\
			&	$	3.321	\pm	0.035	$	&	$	-1.09	\pm	0.12	$	\\
			&	$	3.898	\pm	0.036	$	&	$	-1.21	\pm	0.02	$	\\
			&	$	4.487	\pm	0.032	$	&	$	-1.34	\pm	0.02	$	\\
			&	$	5.101	\pm	0.048	$	&	$	-1.35	\pm	0.02	$	\\
			&	$	5.755	\pm	0.053	$	&	$	-1.33	\pm	0.01	$	\\
			\hline
		\end{tabular}
	\caption{Soret coefficients obtained in this work as a function of salt concentration and temperature. The NEMD simulations were performed using the system equilibrated at an average pressure of 600 bar.
	}\label{tab:Stgromacs1_500}
\end{table}


\begin{table}[h!]
	\begin{center}
		\begin{tabular}{| c  |  c  |  c  |  c |  c  |  c  |}
			\hline
			 $T \; [K]$& $ b\; [ kg \;mol^{-1} ]$ & $S_T \times 10^3  \; [K^{-1}]$ \\
			\hline
			240	&	$	0.938	\pm	0.030	$	&	$	-3.39	\pm	0.24	$	\\
			&	$	2.351	\pm	0.027	$	&	$	-5.79	\pm	0.49	$	\\
			&	$	3.965	\pm	0.076	$	&	$	-3.71	\pm	0.34	$	\\
			\hline
			270	&	$	1.046	\pm	0.03	0	$	&	$	-2.04	\pm	0.22	$	\\
			&	$	2.736	\pm	0.044	$	&	$	-2.31	\pm	0.13	$	\\
			&	$	4.460	\pm	0.032	$	&	$	-2.13	\pm	0.08	$	\\
			\hline
			290	&	$	1.038	\pm	0.022	$	&	$	-0.71	\pm	0.20	$	\\
			&	$	2.728	\pm	0.034	$	&	$	-1.12	\pm	0.11	$	\\
			&	$	4.484	\pm	0.043	$	&	$	-1.50	\pm	0.09	$	\\
			\hline												
		\end{tabular}
	\end{center}
	\caption{Same as Table~\ref{tab:Stgromacs1_500} for average pressure 100~bar.} 
	\label{tab:Stgromacs_100_1}
\end{table}

\begin{table}[h!]
	\begin{center}
		\begin{tabular}{|  c  |  c  |  c  |  c  |  c  |  c  |  c  |  c  |}
			\hline
			 &$ b\; [ kg \;mol^{-1} ]$  & $ N_{pair} $ & $ N_{H_2O} $ & $V~[nm^3]$  &  $ \mu_{id}~[kJ~mol^{-1}]$ & $\mu_{ex}~[kJ~mol^{-1}]$\\
			\hline
Li$^+$ &	$	1.100	$	&	$	16	$	&	$	808	$	&	$	24.20	$	&	$	-6.62	$	&	$	-480.47	\pm	0.45	$	\\
&	$	2.075	$	&	$	31	$	&	$	830	$	&	$	25.53	$	&	$	-5.42	$	&	$	-478.46	\pm	0.38	$	\\
&	$	2.992	$	&	$	43	$	&	$	798	$	&	$	25.09	$	&	$	-4.71	$	&	$	-477.52	\pm	0.65	$	\\
&	$	3.916	$	&	$	54	$	&	$	766	$	&	$	24.69	$	&	$	-4.23	$	&	$	-476.72	\pm	0.57	$	\\
			
\hline					
Cl$^-$ &	$	1.100	$	&	$	16	$	&	$	808	$	&	$	24.27	$	&	$	-6.63	$	&	$	-369.98	\pm	0.85	$	\\
&	$	2.075	$	&	$	31	$	&	$	830	$	&	$	25.56	$	&	$	-5.41	$	&	$	-370.21	\pm	0.96	$	\\
&	$	2.994	$	&	$	43	$	&	$	798	$	&	$	25.18	$	&	$	-4.72	$	&	$	-372.30	\pm	0.87	$	\\
&	$	3.916	$	&	$	54	$	&	$	766	$	&	$	24.77	$	&	$	-4.23	$	&	$	-371.17	\pm	0.96	$	\\

			\hline
		\end{tabular}
	\end{center}
	\caption{Excess chemical potential $\mu_{ex}$ and ideal term $\mu_{id}$ for the Li$^+$ and Cl$^-$ as a function of salt concentration, at 240~K and 600~bar.}\label{tab:free_en-240_1}
\end{table}

\begin{table}[h!]
	\begin{center}
		\begin{tabular}{|  c  |  c  |  c  |  c  |  c  |  c  |  c  |  c  |}
			\hline
			 &$ b\; [ kg \;mol^{-1} ]$  & $ N_{pair} $ & $ N_{H_2O} $ & $V~[nm^3]$  &  $ \mu_{id}~[kJ~mol^{-1}]$ & $\mu_{ex}~[kJ~mol^{-1}]$\\
			\hline
			Li$^+$ &	$	1.100	$	&	$	16	$	&	$	808	$	&	$	24.47	$	&	$	-7.57	$	&	$	-474.61	\pm	0.15	$	\\
&	$	2.075	$	&	$	31	$	&	$	830	$	&	$	25.80	$	&	$	-6.10	$	&	$	-473.02	\pm	0.17	$	\\
&	$	2.994	$	&	$	43	$	&	$	798	$	&	$	25.42	$	&	$	-5.27	$	&	$	-471.75	\pm	0.19	$	\\
&	$	3.916	$	&	$	54	$	&	$	766	$	&	$	25.01	$	&	$	-4.68	$	&	$	-470.69	\pm	0.19	$	\\
			\hline		
			Cl$^-$ &	$	1.100	$	&	$	16	$	&	$	808	$	&	$	24.56	$	&	$	-7.58	$	&	$	-366.82	\pm	0.68	$	\\
&	$	2.075	$	&	$	31	$	&	$	830	$	&	$	25.88 $	&	$	-6.10	$	&	$	-367.43	\pm	0.56	$	\\
&	$	2.994	$	&	$	43	$	&	$	798	$	&	$	25.50	$	&	$	-5.28	$	&	$	-367.61	\pm	0.64	$	\\
&	$	3.916	$	&	$	54	$	&	$	766	$	&	$	25.09	$	&	$	-4.69	$	&	$	-367.80	\pm	0.49	$	\\
			\hline
		\end{tabular}
	\end{center}
	\caption{Same as Table~\ref{tab:free_en-240_1} for 290~K.}\label{tab:free_en-290_1}
\end{table}
\clearpage

\begin{table}[h!]
	\begin{center}
		\begin{tabular}{|  c  |  c   |  c  |  c  |}
			\hline
			$T \; [K]$ & parameter & LiCl & Li$^+$ \\
			 \hline
			240 &	A $ [ mol^{\frac{1}{2}}~ kg^{-\frac{1}{2}}] $	&	$	-21.9118	$	&	$	-15.50	$	\\
&	B	$[ mol^{\frac{1}{2}}~ kg^{-\frac{1}{2}}]$	&	$	0.5663	$	&	$	0.28	$	\\
&	$\beta$	$[ kg~mol^{-1}]$ &	$	-8.9211	$	&	$	-10.50	$	\\
&	C	$[ kg^{2}~ mol^{-2}]$ &	$	3.9684	$	&	$	3.45	$	\\
&	D	$[ kg^{3}~ mol^{-3}]$ &	$	-0.2708	$	&	$	-0.40	$	\\
&	K	$[kJ~ mol^{-1}]$ &	$	-868	$	&	$	-490	$	\\

			\hline
			290 &	A	$ [ mol^{\frac{1}{2}}~kg^{-\frac{1}{2}}] $ &	$	-12.91180	$	&	$	-10.59149	$	\\
&	B	$[ mol^{\frac{1}{2}}~kg^{-\frac{1}{2}}]$ &	$	0.56629	$	&	$	0.64024	$	\\
&	$\beta$	$[ kg~mol^{-1}]$ &	$	-6.92105	$	&	$	-4.64390	$	\\
&	C	$[ kg^{2}~mol^{-2}]$ &	$	2.96837	$	&	$	1.41250	$	\\
&	D	$[ kg^{3}~mol^{-3}]$ &	$	-0.27076	$	&	$	-0.14137	$	\\
&	K	$[kJ~mol^{-1}]$ &	$	-860	$	&	$	-485	$	\\

			\hline
		\end{tabular}
	\end{center}
	\caption{Fitting parameters obtained with 
	equation~(5) in the main text.}\label{tab:parameters}
\end{table}


\section*{Acknowledgments}
We thank the EPSRC (EP/J003859/1)  
for financial support.  We acknowledge the Imperial College High Performance Computing Service for providing computational resources. 

\newpage

\bibliography{biblioFB}
\bibliographystyle{abbrv}

\end{document}